\documentclass[prl,twocolumn,superscriptaddress]{revtex4-2}
\usepackage{amsmath,amssymb,mathrsfs}
\usepackage{graphicx}
\usepackage{colortbl}
\usepackage{braket}
\usepackage{CJK}
\usepackage{bm}
\usepackage{amsfonts}
\usepackage{tabularx}
\usepackage{array}
\usepackage{multirow}
\usepackage{booktabs}
\usepackage[normalem]{ulem}
\begin{document}
%\begin{CJK*}{UTF8}{bsmi}
\title{Many-body computing on Field Programmable Gate Arrays}

\author{Songtai Lv} %(\CJKfamily{gbsn}吕松泰)}
\thanks{These authors contributed equally to this study.}
\affiliation{Key Laboratory of Polar Materials and Devices (MOE), School of Physics and Electronic Science, East China Normal University, Shanghai 200241, China}

\author{Yang Liang} %(\CJKfamily{gbsn}梁洋)}
\thanks{These authors contributed equally to this study.}
\affiliation{Quantum Medical Sensing Laboratory and School of Health Science and Engineering, University of Shanghai for Science and Technology, Shanghai 200093, China}

\author{Yuchen Meng} %(\CJKfamily{gbsn}孟雨晨)}
\affiliation{Key Laboratory of Polar Materials and Devices (MOE), School of Physics and Electronic Science, East China Normal University, Shanghai 200241, China}

\author{Xiaochen Yao} %(\CJKfamily{gbsn}姚晓晨)}
\affiliation{Quantum Medical Sensing Laboratory and School of Health Science and Engineering, University of Shanghai for Science and Technology, Shanghai 200093, China}

\author{Jincheng Xu} %(\CJKfamily{gbsn}徐锦程)}
\affiliation{Quantum Medical Sensing Laboratory and School of Health Science and Engineering, University of Shanghai for Science and Technology, Shanghai 200093, China}

\author{Yang Liu} %(\CJKfamily{gbsn}刘洋)}
\affiliation{Key Laboratory of Polar Materials and Devices (MOE), School of Physics and Electronic Science, East China Normal University, Shanghai 200241, China}

\author{Qibin Zheng} %(\CJKfamily{gbsn}郑其斌)}
\altaffiliation{qbzheng@usst.edu.cn}
\affiliation{Quantum Medical Sensing Laboratory and School of Health Science and Engineering, University of Shanghai for Science and Technology, Shanghai 200093, China}

\author{Haiyuan Zou} %(\CJKfamily{gbsn}邹海源)}
\altaffiliation{hyzou@phy.ecnu.edu.cn}
\affiliation{Key Laboratory of Polar Materials and Devices (MOE), School of Physics and Electronic Science, East China Normal University, Shanghai 200241, China}

\begin{abstract}

A new implementation of many-body calculations is of paramount importance in the field of computational physics. In this study, we leverage the capabilities of Field Programmable Gate Arrays (FPGAs) for conducting quantum many-body calculations. Through the design of appropriate schemes for Monte Carlo and tensor network methods, we effectively utilize the parallel processing capabilities provided by FPGAs. This has resulted in a tenfold speedup compared to CPU-based computation for a Monte Carlo algorithm. By using a supercell structure and simulating the FPGA architecture on a CPU with High-Level Synthesis, we achieve $O(1)$ scaling for the time of one sweep, regardless of the overall system size. We also demonstrate, for the first time, the utilization of FPGA to accelerate a typical tensor network algorithm for many-body ground state calculations. Additionally, we show that the current FPGA computing acceleration is on par with that of multi-threaded GPU parallel processing. Our findings unambiguously highlight the significant advantages of hardware implementation and pave the way for novel approaches to many-body calculations.

%\vspace{0.2cm}
%keywords: Many-body physics, FPGA, Monte Carlo, Tensor networks

\end{abstract}

\maketitle
%\end{CJK*}
\section{Introduction}
Many-body computations are a pivotal research focus within the fields of condensed matter and statistical physics. These problems involve understanding systems composed of a large number of interacting particles, where the collective behavior cannot be easily deduced from the behavior of individual components. The complexity of such systems arises from the exponential growth of the Hilbert space with the number of particles, making exact solutions intractable, even for moderately sized systems. To tackle this challenge, researchers employ a variety of numerical techniques, including sampling methods, such as classical and quantum Monte Carlo (MC)~\cite{Sandvik2010}, as well as blocking methods like Density Matrix Renormalization Group~\cite{DMRG} and tensor networks~\cite{TNreview1,TNreview2}. Among them, tensor networks offer a powerful way to approximate many-body wavefunctions by exploiting the inherent low entanglement in many physical systems, giving them an advantage in handling systems with a sign problem or frustration.

Traditionally, utilizing the central processing unit (CPU) of a general-purpose computer based on the von Neumann architecture, the graphics processing unit (GPU) for parallel processing, or further combining the CPU and GPU in an integrated manner~\cite{DePrince11}, are typical means of implementing many-body calculations. However, certain bottleneck issues, such as the critical slowdown observed in calculations at the vicinity of a phase transition point, persist. Therefore, exploring additional parallel approaches or strategies for many-body computations is imperative to overcome these challenges.

\begin{figure}[t]
\includegraphics[width=0.45\textwidth]{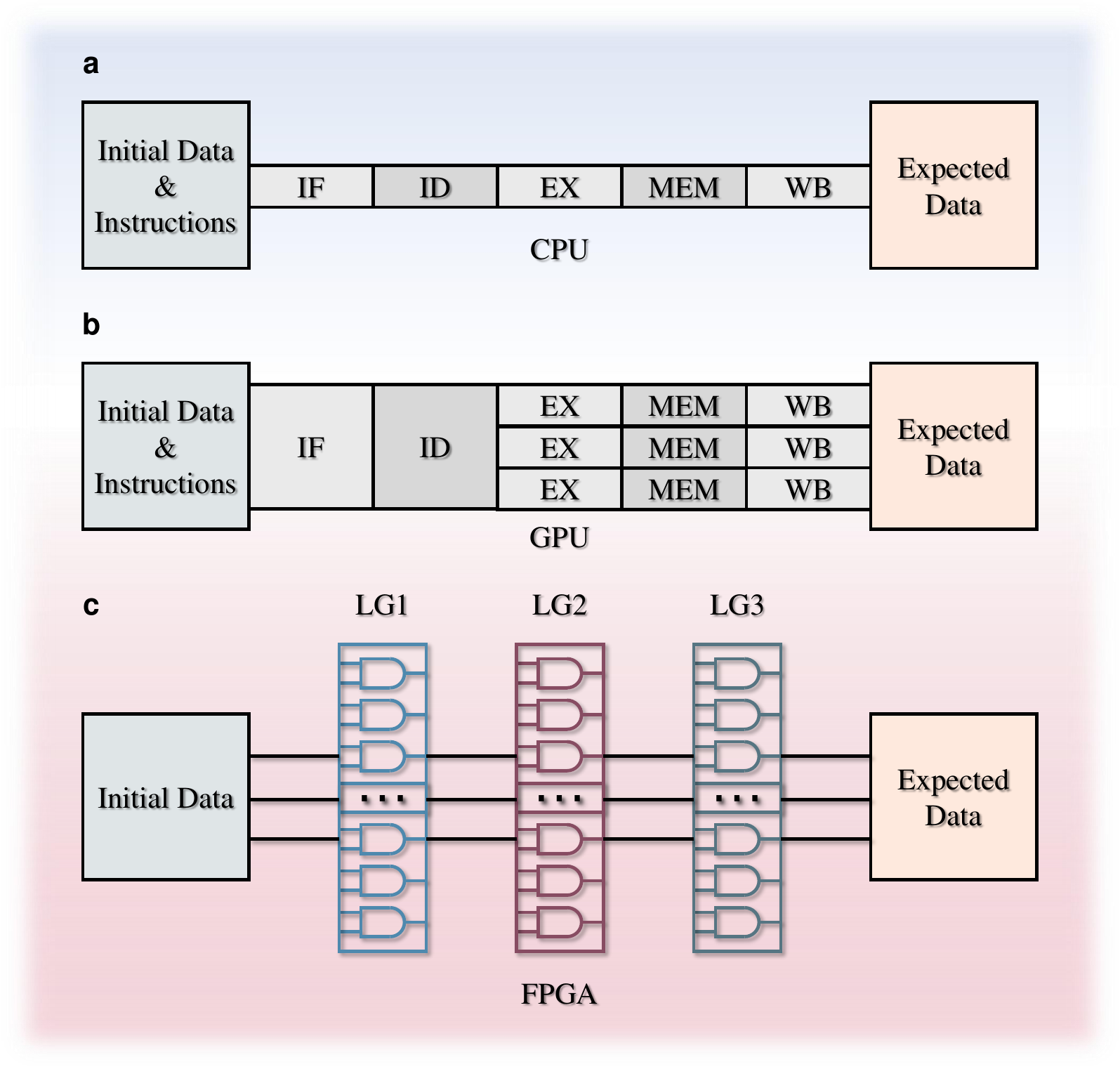}
\caption{\textbf{The architectural differences between FPGA, GPU, and CPU.} \textbf{a}, In the case of CPU, initial data and instructions are stored in the same memory, and after passing through stages of Instruction Fetch (IF), Instruction Decode (ID), Execution (EX), Memory Access (MEM), and Write Back (WB), the expected data is computed. \textbf{b}, In GPU, stages similar to those in CPU are kept, and the decoded instructions are concurrently executed in multiple threads to compute the expected data. \textbf{c}, In FPGA, instructions are implemented by constructing logic gates (LGs), and initial data flows through several LGs to generate the expected data.}
\label{fig:structure}
\end{figure}

Besides from the algorithmic point of view, an alternative approach is to optimize the computations from a hardware perspective. Taking the von Neumann architecture as the first reference example, a typical CPU follows a five-stage instruction execution process consisting of Instruction Fetch (IF), Instruction Decode (ID), Execution (EX), Memory Access (MEM), and Write Back (WB)~\cite{Patterson13}[Fig.~\ref{fig:structure}a]. Moreover, instructions and data are stored in the same memory, which creates a bottleneck~\cite{Backus78} in performance due to the data transfer rate limitation between the processor and memory. Meanwhile, the CPU frequency in the GHz range is difficult to substantially improve due to physical limitations~\cite{Markov14}. These bottlenecks significantly discount the parallel performance of the CPU. For instance, although supercomputers feature multiple parallel processing units, increasing parallelism does not always lead to linear performance growth.

In general, one can improve computational efficiency by modifying or abandoning the von Neumann architecture, but this may come at the cost of losing flexibility. For example, GPUs introduce Single Instruction, Multiple Threads or Single Instruction, Multiple Data execution modes to improving parallelism [Fig.~\ref{fig:structure}b]. However, the execution process still requires the IF, ID, and EX stages. Therefore, we categorize GPUs as hardware approximating the von Neumann architecture. In contrast, an application-specific integrated circuit (ASIC) is designed to perform specific tasks, while losing versatility at the hardware level, but making them highly efficient. For instance, the specialized Tensor Processing Units (TPUs) can significantly accelerate the simulation of quantum many-body dynamic problems~\cite{VidalPRXQuantum2022}. Between the extreme cases of CPU/GPU and ASIC, a Field-Programmable Gate Array (FPGA) can also execute computational tasks in parallel at the hardware level by arranging the computational instructions into logic gates [Fig.~\ref{fig:structure}c]. Take a loop operation for example, FPGAs can utilize pipeline architecture to accelerate the operation by breaking down the computational tasks of each iteration into multiple stages. These stages are executed concurrently in parallel, allowing for overlapping computation and data transfer, which significantly reduces the overall latency. Meanwhile, the structure maintains considerable versatility, as its logic gates are reconfigurable. It is worth mentioning that pipelining and other parallel optimizations can also be applied to the CPU. However, unlike the direct manipulation of logic gates in FPGA, this process still adheres to the von Neumann architecture, and thus, it is not discussed in detail in this paper.

Possessing valuable properties such as high throughput, high energy efficiency, and the capacity to maximize parallelism, FPGAs have emerged as versatile platforms in numerous fields, including quantum chemistry~\cite{RodrguezBorbn2020, Gordon20}, neural networks~\cite{Mittal18}, high-energy physics~\cite{Pandax2023}, etc. For example, in a high-energy physics experiment, FPGAs process vast amounts of data from particle detectors, aiding in the analysis of particle collections and collisions~\cite{Pandax2023}. Interestingly, the structure of the FPGA has even been implemented in a biological system~\cite{Lv2023}. In many-body computations, FPGA has a wide range of applications in MC simulations, for example, to determine the configurations of particle clusters~\cite{Gothandaraman08,Gothandaraman09, Gothandaraman10, Gothandaraman11, Cardamone19} or to simulate the integer-variable spin models like the various Ising models~\cite{FPGA_ising,Yoshimura2016,BaityJesi2013,spinglass2020, Okuyama17, Waidyasooriya18, Liu19, Waidyasooriya19, Liu21, Waidyasooriya21, Chung22, Kubuta23}. Although FPGAs have been proposed for handling tensor structures involved in neural networks~\cite{Zhang21} and quantum circuits~\cite{Levental21}, their substantial parallel advantages have not been fully exploited in the realm of many-body systems with more degrees of freedom, particularly in the area of tensor networks algorithms.

Here, we implement two typical many-body algorithms on FPGA, utilizing the two-dimensional (2D) XY model and the one-dimensional (1D) Heisenberg chain as examples. Using only a single FPGA chip, we showcase its significant parallel acceleration capabilities (The details of the chip are shown in the Methods Section). First, we construct a parallel scheme for the Metropolis MC updating procedure of the two-dimensional classical XY model, achieving a tenfold speedup compared to the CPU. Second, we restructure the Infinite Time-Evolving Block Decimation (iTEBD)~\cite{TEBD}, a typical tensor network algorithm, making it suitable for hardware execution and also beating the CPU speed for the first time. Additionally, we demonstrate that the current acceleration achieved by FPGA computing is comparable to that of multi-threaded GPU parallel computation. Our work has achieved hardware FPGA acceleration for many-body computations, offering a novel approach to this field in general.

It is worth noting that utilizing new hardware like FPGA for computations does not diminish the importance of algorithms. On the contrary, algorithms remain crucial, and it is essential to modify algorithms suitable for parallel execution on FPGA to fully leverage its parallel processing capabilities. For example, in Reference~\cite{Nikhar24}, higher-order probabilistic bit (p-bit) Ising machines are used to achieve $O(1)$ scaling for one MC sweep in dense graph problems, independent of the total system size. Next, we modify algorithms suitable for parallel execution on FPGA for Metropolis Monte Carlo and iTEBD computations, using the 2D XY model and the 1D Heisenberg chain as examples, respectively.

\begin{figure}[t]
\includegraphics[width=0.5\textwidth]{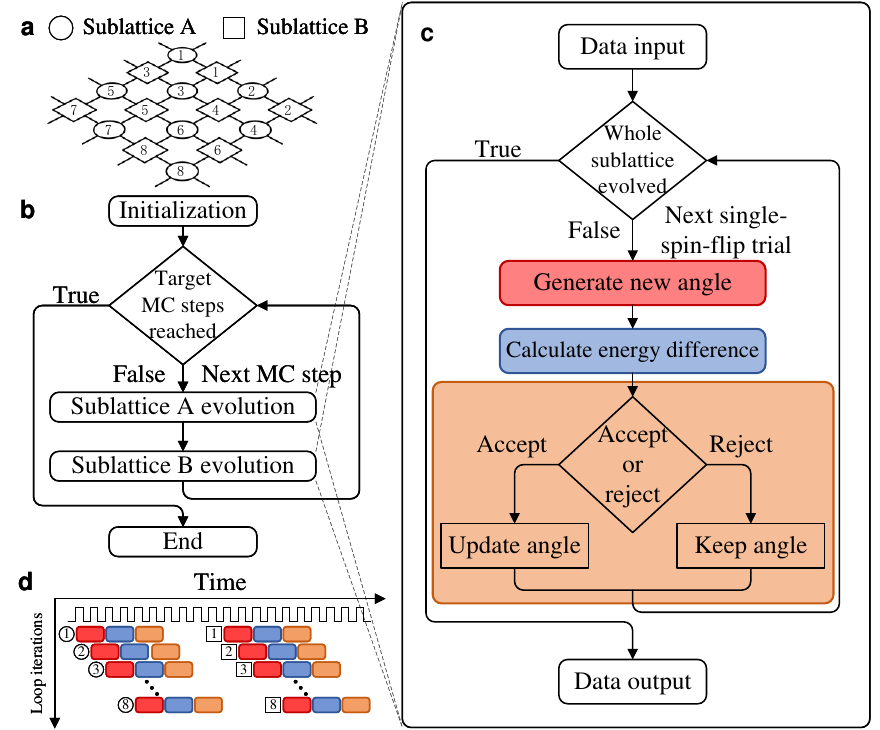}
\caption{\textbf{ Schematic diagram of the MC simulation for the XY model on the square lattice}. \textbf{a}, The XY model is decomposed into two sublattices A and B. \textbf{b}, Flowchart of the MC simulation where A and B are alternately evolved. \textbf{c}, All the spins in the sublattice are updated by the single-spin-flip trials (SSFTs), where each trial is split to three small stages. \textbf{d}, Pipelined time series sketch of one MC step. The colored rectangles represent three stages of SSFT, and the numbered circles and squares represent which spin is updated. The SSFTs is tailored into FPGA in the form of pipelined small stages.}
\label{fig:mctric}
\end{figure}

\section{Results}
\subsection{Parallel design of MC simulation for the XY model}

We first focus on the 2D classical XY model, which is renowned for the Kosterlitz-Thouless phase transition~\cite{Kosterlitz1973} and is described by the following Hamiltonian:
\begin{equation}
    H_{\rm XY} =  - \sum_{ (\bm{x}, \bm{\delta}) } {\rm cos}\left(\theta_{ \bm{x} } - \theta_{ \bm{x} + \bm{\delta} }\right)
    \label{eq.XYHamiltonian}
\end{equation}
where $\bm{x}$ denotes a site on the square lattice, $\bm{\delta}$ is a unit vector connecting this site with one of its nearest neighbours and the variable $\theta_{ \bm{x} } \in [0, 2\pi)$ represents the direction of the spin at site $\bm{x}$.

We employ the straightforward Metropolis algorithm to evolve the configurations of the system at a temperature $T = 0.85$, which is close to the KT phase transition. Due to bipartition of the square lattice, (In other words, each unit cell contains two lattice sites,) half of the spins can be evolved independently and in parallel. Therefore, we divide the lattice system into two sets of subsystems and update them alternately(as illustrated in Fig.~\ref{fig:mctric}) by following the approach used in Ref.~\cite{FPGA_ising}.
In each subsystem, the MC procedure is further divided into three steps. First, random parameters are generated for each spin as the initialization of the Metropolis process. Second, the local internal energy difference caused by spin flipping is calculated, and the probability of flipping is determined. Third, a probabilistic decision to finish the Metropolis process is made, and the spin states are updated. More details of the Metropolis process are provided in the Supplementary Section.

During the evolution process for each subsystem, the iteratives steps for each spin are independent of each other, thereby avoiding the serial iteration of independent steps and fully utilizing available hardware resources for acceleration. The computation rate can be further improved by increasing the number of lattice sites in the unit cell. After implementing the entire MC algorithm into the FPGA hardware, we can readily employ the pipeline architecture to accelerate the computation.
We calculate the energy of the system and obtain results consistent with those from CPUs (details shown in the Supplementary Section). The computational speed advantage of FPGA compared to CPU and GPU is demonstrated in the Speedup Results part of this section.

%%%%%%%%%%%%%%%%%%%%%%%%%%%%%%%%%%%%%%%%%%%%%%%%%
\subsection{Parallel design of iTEBD for the Heisenberg chain}

The ground state $\ket{\Psi}$  of a given system with the Hamiltonian $H$ can be effectively determined by evolving an appropriate initial state $\ket{\Psi_0}$ along the imaginary-time path, expressed as $\ket{\Psi} = \lim_{\tau\to\infty} \exp(-\tau H) \ket{\Psi_0}$. In a tensor network calculation, discretizing the evolution operator $\exp(-\tau H)$ into tensors greatly enhances the ability to capture entanglement in quantum systems, with the iTEBD algorithm serving as a prominent example of this approach. We then demonstrate the implementation of the iTEBD algorithm on FPGA for the 1D antiferromagnetic Heisenberg model. The Hamiltonian for this model is as follows:
\begin{equation}
    H_{\rm H}=\sum_{\langle ij\rangle}H_{ij}=\sum_{\langle ij\rangle}S^x_iS^x_j+S^y_iS^y_j+S^z_iS^z_j,
    \label{eq:H_Hei}
\end{equation}
where the interaction $H_{ij}$ is applied locally between the nearest neighbor sites $i$ and $j$, and $S^\mu_i = \sigma^\mu/2$, with $\sigma^\mu$ being the three Pauli matrices ($\mu=x,y,z$).

In the usual iTEBD setup, the initial wavefunction $\ket{\Psi}$ of the system is constructed as $\ket{\Psi}= \cdots\lambda_{2,ij}A^\alpha_{jk}\lambda_{1,kl}B^\beta_{lm}\lambda_{2,mp}\cdots\ket{\cdots\alpha\beta\cdots}$, where $A$,$B$,and $\lambda_{1,2}$ are local tensors, and all the virtual indices $i,j$, etc are contracted. These ingredients are updated by local imaginary time evolution $\ket{\tilde{\Psi}}=\prod_{\langle ij\rangle}U_{T,ij}\ket{\Psi}=\prod_{\langle ij\rangle}e^{-\tau H_{ij}}\ket{\Psi}$ until convergence, where the local gate $U_T$ has a matrix representation
\begin{equation}
    U_T = 
    \begin{pmatrix}
        e_{0} & 0        & 0        & 0 \\
        0        & e_{1} & e_{2} & 0 \\
        0        & e_{2} & e_{1} & 0 \\
        0        & 0       & 0        & e_{0}
    \end{pmatrix}
    \label{eq.Umatrix}
\end{equation}
with the bases $\{\ket{\uparrow\uparrow},\ket{\uparrow\downarrow},\ket{\downarrow\uparrow},\ket{\downarrow\downarrow}\}$, and the nonzero elements of $U_T$ are $e_{0} = \bra{\uparrow\uparrow} e^{-\tau H_{ij}} \ket{\uparrow\uparrow} = \bra{\downarrow\downarrow} e^{-\tau H_{ij}} \ket{\downarrow\downarrow}$, $e_{1} = \bra{\uparrow\downarrow} e^{-\tau H_{ij}} \ket{\uparrow\downarrow}=\bra{\downarrow\uparrow} e^{-\tau H_{ij}} \ket{\downarrow\uparrow}$, $e_{2} = \bra{\uparrow\downarrow} e^{-\tau H_{ij}} \ket{\downarrow\uparrow}=\bra{\downarrow\uparrow} e^{-\tau H_{ij}} \ket{\uparrow\downarrow}$. In our calculation, the length of one time step $\tau$ is set as $\tau = 0.01$. Moreover, in the evolution process, it is necessary to utilize singular value decomposition (SVD) approximation to ensure that the virtual bond dimension is fixed as a finite value $D_b$.  

\begin{figure}[t]
\includegraphics[width=0.5\textwidth]{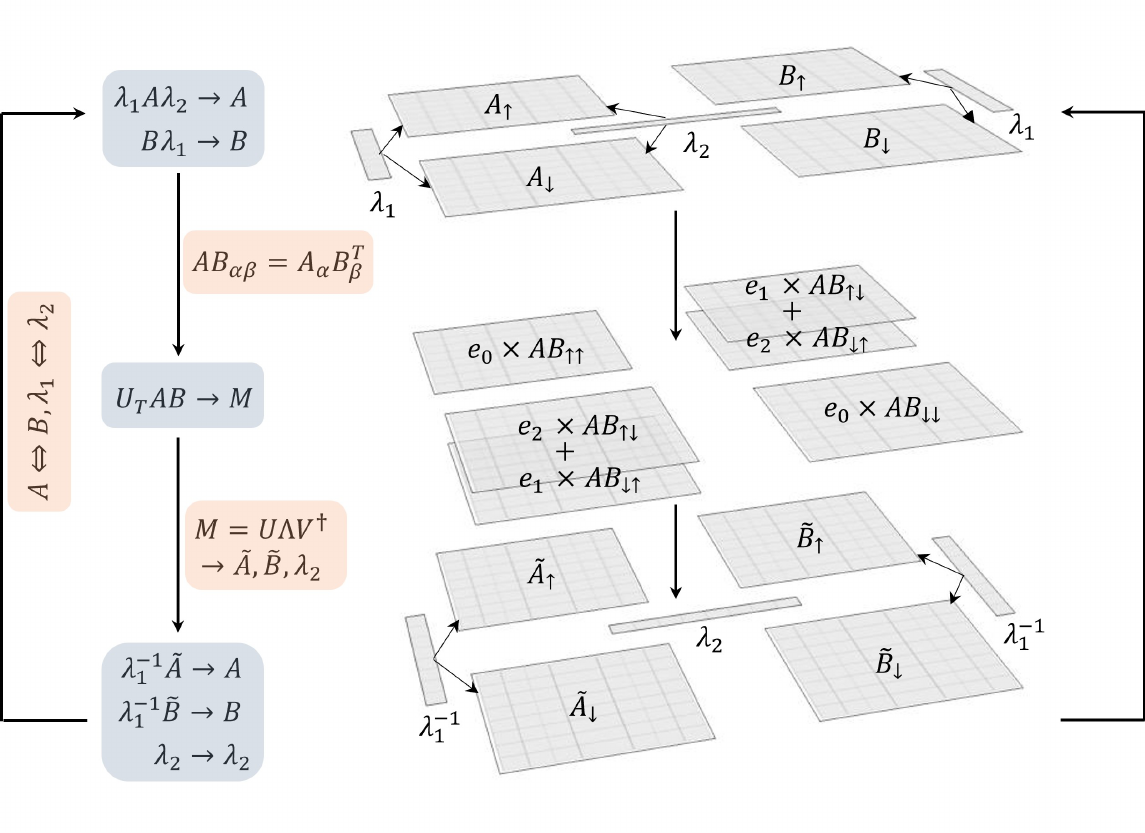}
\caption{\textbf{Schematic diagram of the iTEBD calculation for the Heisenberg model.} The left side depicts the variable flowchart, while the right side shows the corresponding matrix representations. The iTEBD process consists of three steps:
        First, after multiplying $\lambda_1$ and $\lambda_2$ into $A_{\uparrow}$, $A_{\downarrow}$, $B_{\uparrow}$, and $B_{\downarrow}$, $A$ and $B$ are directly multiplied to obtain the matrix. Second, $AB$ undergoes evolution through gate $U_T$ to yield $M$, and SVD is applied to $M$ to obtain matrices $U$, $\Lambda$, and $V$, which are truncated to dimension $D_b$ to generate $\tilde{A}$, $\tilde{B}$, and $\lambda_2$. Third, $\lambda_1^{-1}$ is multiplied into $\tilde{A}$, $\tilde{B}$ to obtain the new $A$ and $B$ for the next iteration of the iTEBD process.}
\label{fig:iTEBD}
\end{figure}

When performing tensor network computations on CPUs, tensors or matrices are often defined as classes. This allows for intuitive and crucial operations like permutation and reshaping of the classes themselves when contracting tensors with different indices. These operations enhance the intuitiveness and efficiency of tensor operations. However, in the existing FPGA architecture, the input and output data ultimately manifest in the tensor elements themselves. Operations like permutation and reshaping of tensors, which are suitable for CPUs, instead significantly slow down computation speed and increase unnecessary resource consumption on FPGAs. Therefore, in FPGA implementation, one needs to modify the iTEBD algorithm from the perspective of tensor elements rather than classes. Following this principle, we redesign the iTEBD algorithm suitable for parallel execution on FPGAs (Fig.~\ref{fig:iTEBD}), avoiding excessive tensor operations. (Note that this improvement is also applicable to higher-order tensors that appear in high-dimensional tensor network algorithms.) For instance, instead of relying on the third-order tensor $A_{ij}^\alpha$, we employ two matrices $A_\uparrow$ and $A_\downarrow$, significantly simplifying the process and enabling parallelism. A detailed discussion on the equivalence of this new form with the original iTEBD can be found in the Supplementary Section.

In the iTEBD process, separated by the SVD step, the entire procedure is structured into three stages (Fig.~\ref{fig:iTEBD}): pre-SVD, SVD, and post-SVD. In the pre-SVD stage, $\lambda$s are multiplied into matrices $A_\uparrow$,  $A_\downarrow$, $B_\uparrow$ and $B_\downarrow$, followed by the multiplication of $A$s and $B$s to derive a new matrix $AB$ by $AB_{\alpha\beta} = A_\alpha B_\beta^{\rm T}$  ($\alpha$, $\beta$ can be $\uparrow$ or $\downarrow$). Subsequently, the non-zero matrix elements of $U$ are used to reassemble $AB$. In the SVD stage, the reassembled $AB$ undergoes SVD using a parallelizable two-sided Jacobi algorithm (detail in the Supplementary Section). In the post-SVD stage, the obtained results are truncated to preserve the dimension $D_b$, resulting in new matrices $A$, $B$ and $\lambda$s. These newly derived matrices are reintegrated into the initial step, facilitating the iterative process until convergence. Using the wavefunction from these processes, we get convergent ground state energy consistent with the exact solution for Heisenberg chain (details shown in the Supplementary Section). The computational speed advantage of FPGA compared to CPU and GPU is shown in the next section.

\subsection{Speedup results}

We compare the computational time results on CPU, GPU and FPGA for both MC simulation for the XY model and iTEBD calculation for the Heisenberg model, showcasing the significant acceleration achieved by FPGA. The same set of modified MC and iTEBD algorithms are deployed across four different platforms: CPU, FPGA in sequential style, and FPGA in pipelined parallel style, GPU with 32 threads per warp. For the first two platforms, the simulations were designed in a completely sequential manner. We then measured the computation time of one MC step and one iTEBD step for all cases.

\begin{widetext}

\begin{figure}[t]
\includegraphics[width=0.9\textwidth]{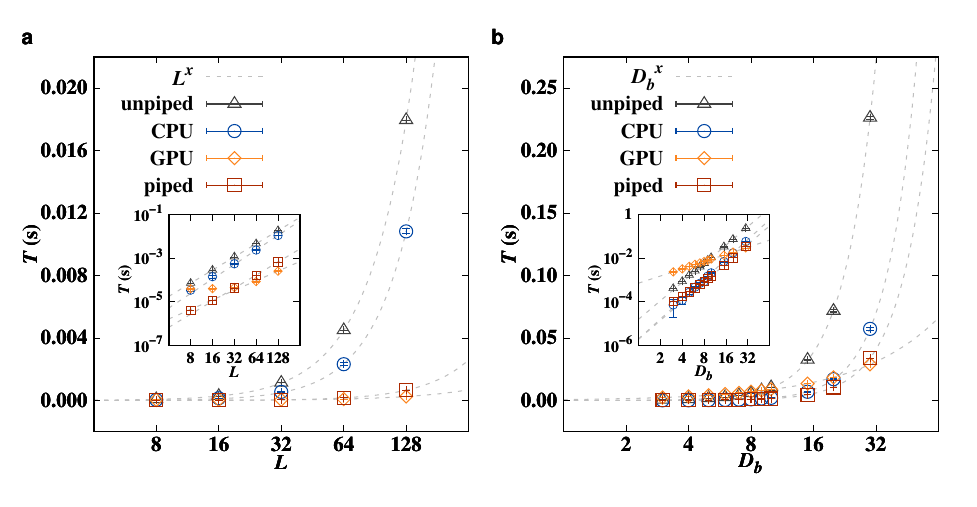}
\caption{ \textbf{The computational time per step for a, MC simulation of the XY model and b, iTEBD calculation of the Heisenberg chain.} The red squares, black triangles, orange diamonds, and blue circles represent the computational time for FPGA in pipelined parallel style, FPGA in sequential style (unpiped), GPU, and CPU, respectively. The error bars indicate a 95\% confidence interval. The gray dashed lines represent the fitted results of the computational time with the fitting function $X^x$, where $x$ is the fitting parameters, and $X = L$ or $D_b$ for \textbf{a} or \textbf{b}. In \textbf{a}, the fitted results for $x$ for FPGA in parallel style, FPGA in sequential style, GPU, and CPU are 1.99, 2.00, 1.46, and 2.20, respectively. In \textbf{b}, the corresponding values of $x$ are 2.88, 2.82, 1.09, and 3.02. The insets illustrate the corresponding data in log-log scale.}
\label{fig:Time1}
\end{figure}

\end{widetext}

In the MC simulations for the XY model, we executed 10,000 MC steps for five different sizes $L =$ 8, 16, 32, 64, and 128 on each platform to measure the average computation time. The results are displayed in Fig.~\ref{fig:Time1}a. While the computational speed of the FPGA in sequential style is slower than that of the CPU, significant speed improvement is achieved with the FPGA in pipelined parallel style. For example, at the system size $L=128$, the computation speed of the FPGA in parallel is 16.5 times that of the CPU, 27.3 times that of the FPGA in sequential style. For these three platforms, the computation time is approximately proportional to $L^2$, with slight variations. We also optimize GPU performance with 32 threads per warp. The results show that FPGA parallelization achieves better acceleration for $L\le 32$. However, since only pipeline parallelism is used on the FPGA, its performance becomes slightly less efficient than GPU for larger $L$, where multi-threading on the GPU is effectively equivalent to unrolling. Additionally, the multi-threaded GPU acceleration generally outperforms the multi-core CPU. We confirmed this fact using OpenMP results with 2 to 32 CPU cores and demonstrated that the acceleration on the CPU is far from linear with respect to the number of cores (details in the Supplementary Section).  Compared to the higher acceleration achieved with the proposed probabilistic computer on the Ising model, the relatively lower acceleration 
in the XY model is mainly due to the use of higher precision variables (64-bit floating-point) instead of the integer variables in the Ising model. This consumes more hardware resources, limiting our ability to further optimize using the unroll approach.

Figure~\ref{fig:Time1}b illustrates the computation time of one iTEBD step for the Heisenberg chain system. Starting from several fixed initial states, we executed 10,000 to 100,000 iteration steps for 11 different bond dimensions $D_b \le 30$, aiming to achieve convergent final states with the same level of precision. Similar to the MC simulation, the FPGA in sequential calculations takes the longest time compared to the other three platforms, while the computational speed of FPGA in parallel surpasses that of the CPU. However, the acceleration effect is not as pronounced as that of the MC calculations. For instance, at $D_b = 30$, the computation speed of the FPGA in parallel is 1.7 times that of the CPU, 6.7 times that of the FPGA in sequential style, and also achieves better acceleration than GPU with 32 threads per warp for $D_b < 30$. Nevertheless, this result remains encouraging because only the pipeline parallelism in FPGA is used so far. This suggests the potential for achieving more pronounced acceleration with larger $D_b$.

From Fig.~\ref{fig:Time1}, it can be observed that the error in the runtime of the FPGA in parallel is smaller compared to the CPU for both the MC and the iTEBD calculation. This highlights another advantage of FPGA: deterministic latency. This is attributed to the architectural differences between the two platforms. When executing complex tasks on the CPU, rescheduling, instruction fetching, decoding, and other operations introduce variability in the runtime for each execution. However, due to its hardware programmability, each operation in an FPGA is fixed within a clock cycle, resulting in smaller runtime errors compared to the CPU.

Although in both the MC and iTEBD cases, the CPU outperform the FPGA in sequential due to the CPU's higher frequency and the relative simplicity of the algorithm currently being considered, the parallelization effect on an FPGA generally results in higher computational efficiency than the CPU. Compared to the parallelism of CPUs, which mainly relies on multi-core processors and faces issues such as communication and synchronization overheads, as well as shared memory access conflicts, the parallelization on FPGA is much more efficient. FPGAs enable deterministic memory access by allowing direct data transfer between logic blocks without intermediate caching. This reduces overheads and simplifies processes. Additionally, parallel access to multiple memory blocks boosts bandwidth and avoids conflicts. These suggest that appropriately increasing the complexity of 
\begin{widetext}

\begin{figure}[t]
\includegraphics[width=0.9\textwidth]{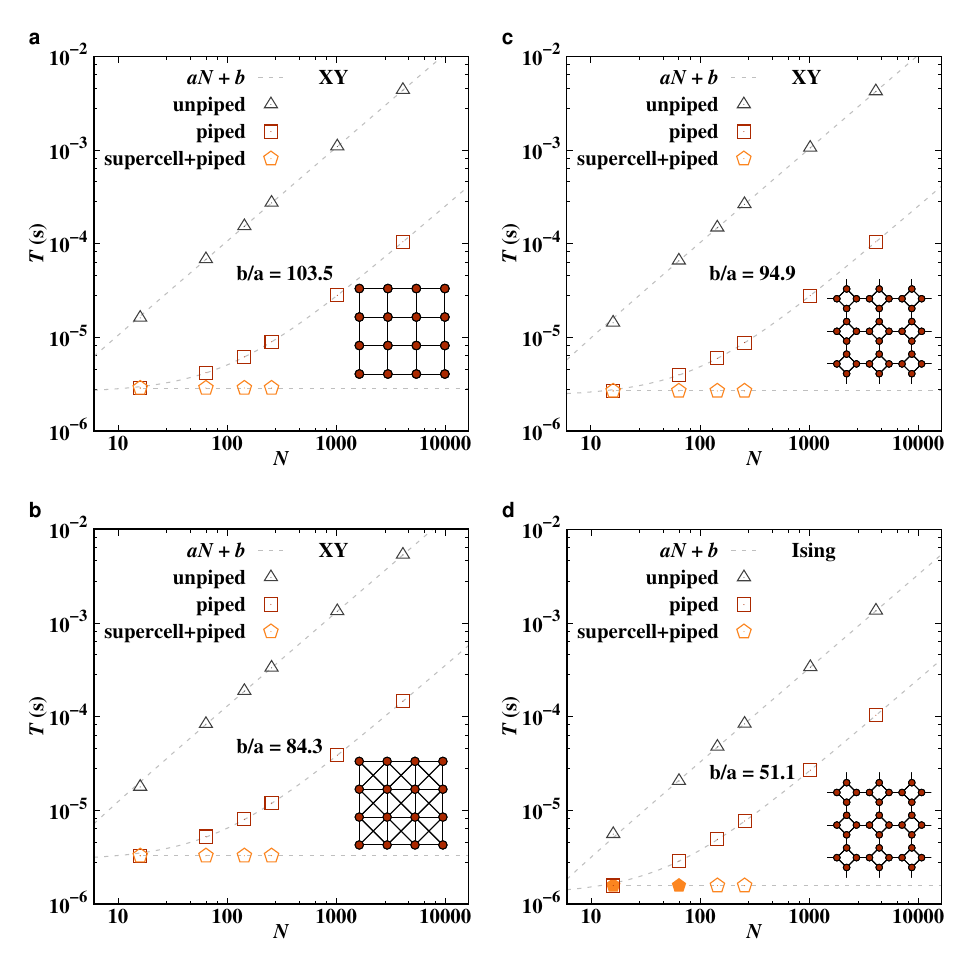}
\caption{\textbf{The computational time (simulated on a CPU for FPGA) per MC sweep for a, the XY model on square lattice, b, the XY model with NNN interactions on square lattice, c, the XY model on square-octagonal lattice and d, the Ising model on square-octagonal lattice, with the system lattice sizes $N$ = 16, 64, 144, 256, 1024, and 4096.} The black triangles, red squares, and orange pentagons represent the computational time for FPGA in sequential style, in pipelined parallel style with 4 sites in one unit cell, and pipelined parallel style with a supercell, respectively. In \textbf{d}, the cases $N$ = 16 and 64 were verified on our FPGA chip, labeled as solid pentagons. The gray dashed lines represent the fitted results with $aN + b$, where $N$ is the number of spins. The ratios $b/a$ extracted from the results of the pipelined parallel style with 4 sites in one unit cell, are shown at the center of each panel. The insets in the bottom-right corner illustrate the lattice configurations. }
\label{fig:O1time}
\end{figure}

\end{widetext} 
the existing algorithm to align with the FPGA architecture can help improve the computation rate. Here, we take MC calculations as an example, increasing the number of sites in the unit cell during the computation to compare with the previous results of bipartite lattice calculations.

We first set the number of lattice sites in the unit cell to four and calculate the XY model on a square lattice with only nearest-neighbor (NN) and with both NN and next-nearest-neighbor (NNN) interactions, as well as the XY model and Ising model on an octagon-square lattice with NN interactions, shown in Fig. 5. (Note that, except for $N$ = 16 and 64 in Fig.~\ref{fig:O1time}d, which were emulated on our chip, all other data were simulated on a CPU following the FPGA architecture). Since the time per sweep on the FPGA scales linearly with the number of lattice sites $N$ (as confirmed in Fig.~\ref{fig:Time1}a), we fit the relationship between time and $N$ using $aN + b$. For the unpiped results, $b \ll a$, whereas for the piped results, $b/a$ is significantly larger, which can be used as a metric to characterize the acceleration effect. The value of $b/a$ shown in Fig.~\ref{fig:O1time}a is larger than that from Fig.~\ref{fig:Time1} ($b/a = 35.0$), indicating that increasing the number of lattice sites in a unit cell for the same system indeed contributes to improving the computation speed. The pipelined results in Figs.~\ref{fig:O1time}b and \ref{fig:O1time}c, which consider NNN interactions and different lattice systems, exhibit similar acceleration effects. In Fig.~\ref{fig:O1time}d, the results for the Ising model are also show a speedup compared to the CPU results in reference~\cite{Chowdhury2023}, suggesting that our system can also be extended to use p-bits in reference~\cite{Chowdhury2023} for simulating quantum models. Furthermore, we consider a supercell by setting the number of lattice sites in the unit cell to $N/4$, and with pipelining, it achieves $O(1)$ scaling of computation speed with the number of lattice sites. In fact, for periodic many-body models with interactions beyond NN or NNN, a similar acceleration can be achieved by increasing the number of lattice sites in the unit cell. For irregular graphs, more sophisticated algorithms are needed to sparsify the graph~\cite{Nikhar24}. 
These optimizations require converting the existing variables into more hardware components. Although this makes the program appear lengthy, it can be easily achieved by appropriately encoding and sorting each variable that describes the hardware to generate the code needed for execution.

%%%%%%%%%%%
\section{Discussion}

While the ultimate solution to the exponential wall problem in many-body systems will eventually rely on quantum computers~\cite{ibm2023}, the full utilization of classical algorithms and hardware remains meaningful at present. Generally, the time complexity for computing many-body systems follows a growth pattern of $hX^a$, where $X$ represents the controllable degrees of freedom for the particular problem, such as system size $L$ in Monte Carlo simulations or bond dimension $D_b$ in tensor network calculations. Improvements in algorithms are reflected in the reduction of $a$, while hardware parallel optimization alters $h$. Therefore, comprehensive improvements in both algorithms and hardware will contribute to the efficiency of many-body computations. 

From an algorithmic perspective, whether it's MC or iTEBD, our method is universal and can be naturally extended to solve other many-body models. Additionally, for MC algorithms, we can further improve the computational efficiency for MC simulation by applying cluster algorithms like Swendsen-Wang algorithm, and extend the system to general and irregular lattices by applying graph coloring strategy~\cite{Aadit22}. Back to the hardware optimization, we can further accelerate the MC simulation by utilizing a more suitable random number generator for FPGA~\cite{Lin18} and optimizing the operand bit width with strategy like mixed precision~\cite{Chow12}.

Through comparison with a 32-thread GPU, we found that for the current relatively small $L$ and $D_b$, parallel FPGA acceleration has a slight advantage. However, the GPU demonstrates better scaling behavior with increasing $L$ or $D_b$. This indicates that in cases of larger $L$ or $D_b$, the parallel FPGA acceleration scheme needs improvement. The GPU's advantage comes from having more memory, allowing more threads to achieve better acceleration (details in the Supplementary Section). Although the GPU's multi-threading approach is similar to unrolling, and its current computational performance scales well with the number of systems, the acceleration achieved by the GPU becomes less pronounced as the number of threads increases. In contrast, the FPGA demonstrates a more substantial acceleration effect when transitioning from serial to pipelined execution. Moreover, increasing the number of lattice sites in the unit cell further amplifies the acceleration effect, highlighting the significant potential of FPGA for parallel acceleration when more hardware resources are available. Additionally, the implementation of many-body computations across different hardware platforms is not a matter of competition. On the contrary, their synergy and complementarity can better drive breakthroughs in overcoming computational bottlenecks in the future. 

For the case of iTEBD calculations in this paper, our primary goal is to implement a basic tensor network parallel algorithm on FPGA, thereby providing a reference for the hardware implementation of more complex tensor network parallel algorithms~\cite{Secular20}. Therefore, we have focused on avoiding tensor reshaping and permutation operations at the algorithmic level and have only employed pipeline parallelism at the hardware level. However, the most time-consuming aspect of tensor calculations is the SVD process. With improvements in hardware programming for the SVD process, such as utilizing modified systolic arrays~\cite{Luk1985}, coordinate rotation digital computer (CORDIC)-like technology~\cite{Zhang2017}, and probabilistic-bit implementations~\cite{Chowdhury2023, Niazi24, Singh24}, the computational speed of tensor network algorithms on FPGA will be significantly enhanced.

Another aspect worth discussing is the current limitations of resources on FPGA chips. For example, although the Block Random Access Memory (BRAM) of the chip we are using is only 15.6Mb, it can handle many-body problems of moderate scale for $L$ or $D_b$. However, in tensor network computations, models involving sign problems~\cite{Liu2023} and the competition of various phases~\cite{Liu2022,Zou2023} require larger $D_b$, and solving these problems requires FPGAs with larger memory. The success of TPU~\cite{VidalPRXQuantum2022, Vidal23} on tensor network calculations demonstrates that this is not a technically insurmountable problem. Therefore, optimizing multi-body computations on FPGAs can stimulate the development of chip hardware.

In summary, we have employed FPGA to achieve hardware acceleration for two distinct many-body computation methods. In the MC evolution of the 2D classical XY model, subsystem parallelization significantly boosts the evolution speed, crucial for mitigating issues related to critical slowdown. In the iTEBD tensor network calculation of the 1D Heisenberg chain, we have devised a structure optimized for hardware dataflow, streamlining computational steps based on tensor elements. This structural optimization, combined with parallelization, results in speeds surpassing those attainable on a CPU. Notably, this represents the first FPGA implementation of a tensor network algorithm. Our current FPGA acceleration matches the performance of multi-threaded GPU parallel processing. Our findings offer a novel perspective on many-body computations, fostering interdisciplinary collaboration between the fields of many-body algorithms and hardware.

\section{Methods}
%\subsection{FPGA hardware}

{\it FPGA hardware }The FPGA chip model used in this work is XC7K325TFFG900-2. The driving pulse period is 10ns, corresponding to a frequency of 0.1GHz. This chip has 890 Block RAM (BRAM) units, each with a capacity of 18 kilobits (Kb), abbreviated as BRAM$\_$18K. In other words, the entire chip has only 15.6 Mb of memory available for storing data, which severely limits the size of the computational system. Additionally, the chip features 840 Digital Signal Processing slices of the 48E version (DSP48E), 407,600 flip-flops (FF), and 203,800 look-up tables (LUT).

In the MC simulation for the XY model with $L$=128, the utilization percentages of BRAM$\_$18K, DSP48E, FF, and LUT are 42$\%$, 39$\%$, 5$\%$, and 18$\%$, respectively. In the iTEBD calculation for the Heisenberg chain with $D_b$=30, the utilization percentages of BRAM$\_$18K, DSP48E, FF, and LUT are 42$\%$, 63$\%$, 24$\%$, and 69$\%$, respectively. For Fig.~5, the CPU simulation following the FPGA architecture is performed using the High-Level Synthesis (HLS) strategy with the Xilinx Vivado HLS software.

%\subsection{CPU hardware}

{\it CPU hardware }The CPU used in this work is the Intel Xeon Gold 6230, with a clock frequency of 2.10 GHz. The CPU computations are performed in one core of the Dell PowerEdge R740 server. In a single-core environment, the peak memory for the MC simulation of $L = 128$ XY system can reach 80 Mb, with an average memory of 24 Mb. For the iTEBD calculation of a $D_b = 30$ Heisenberg system, both the peak and average memory are 16 Mb.

%\subsection{GPU hardware}

{\it GPU hardware }The GPU used in this work is the Quadro K620, with a clock frequency of 1.058 GHz. For both of the $L = 128$ XY system and $D_b = 30$ Heisenberg system, the memory usage, including the data memory and CUDA library cache, is 168Mb.

\subsection{Acknowledgments}

We thank Jianguo Ma, Deyan Sun for helpful discussions. This work is supported by National Natural Science Foundation of China Grant (No. 12274126, 12105177) and National Key R\&D Program of China (No. 2023YFF0719200).

%\bibliography{FPGA}
%

\pagebreak

\newpage

\widetext
\begin{center}
\textbf{\large Supplementary information for \\``Many-body computing on Field Programmable Gate Arrays"}
\end{center}

\setcounter{equation}{0}
\setcounter{figure}{0}
\setcounter{table}{0}
\makeatletter
\renewcommand{\thefigure}{S\arabic{figure}}
\renewcommand{\thetable}{S\arabic{table}}
\renewcommand{\theequation}{S\arabic{equation}}
\renewcommand{\bibnumfmt}[1]{[S#1]}
\renewcommand{\citenumfont}[1]{S#1}
\makeatother
\newcolumntype{C}{>{\centering\arraybackslash}X}

%\maketitle
%\end{CJK*}
%\newpage
\section{Detail of the MC calculation for the XY model}
As indicated by the Hamiltonian in Eq.~(1) in the main text, the 2D classical XY model we studied only has nearest neighbour interaction, and the direction of each spin is confined in $[0, 2\pi)$. We use periodic boundary conditions in the calculation. In step 1, a new angle $\theta'_{\bm{x}}$ and a new probabilistic decision factor $p_{\bm{x}}$ are generated for each spin, where $\theta'_{\bm{x}}$ is sampled from a uniform distribution in $[0, 2\pi)$, and $p_{\bm{x}}$ is sampled from a uniform distribution in $[0, 1)$.
The random number generator is constructed using the linear congruence method, whose recurrence relation is $X_{n+1}^{\bm{x}} =[~(a X_n^{\bm{x}} + c) ~ {\rm{mod}}~m~]$, where $X_{n+1}^{\bm{x}}$ and $X_n^{\bm{x}}$ are the pseudo-random values of the $(n+1)$-th and $n$-th iterations for the spin at site $\bm{x}$, respectively, and $a = 25214903917$, $c = 11$, $m = 2^{48}$.
Each spin is initialized with its own unique random seed $X_{0}^{\bm{x}}$, enabling each spin to undergo an independent Metropolis process. 
The sequence of pseudo-random values is designed as 64-bit non-negative integers, which are stored in dedicated registers and accessed directly via the FPGA’s fabric.
With a constant bandwidth of 64 bits per 10 nanoseconds, generating one random number incurs a latency of 20 nanoseconds. This latency is longer than the approximately 10 nanoseconds time required for the CPU to generate a random number. Note that the CPU runtime results are the averaged results obtained after running $10^{10}$ operations. 
The newly generated random numbers are then divided by the modulus $m$ or $m/2\pi$ to obtain the probabilistic decision factor $p_{\bm{x}}$ or the angle $\theta'_{\bm{x}}$, respectively.
In step 2, the probability that the spin $S_{\bm{x}}$ points in a new direction was calculated as $P(\Delta E_{\bm{x}}) = e^{- \beta \Delta E_{\bm{x}}}$, where $\beta$ is the inverse temperature $\frac{1}{T}$, and $\Delta E_{\bm{x}}$ is the local internal energy difference caused by the spin rotation,
\begin{equation}
    \Delta E_{\bm{x}} = 
        - \sum_{ \bm{\delta} } \left[
            {\rm cos}\left(\theta'_{ \bm{x} } - \theta_{ \bm{x} + \bm{\delta} }\right)
            + {\rm cos}\left(\theta_{ \bm{x} } - \theta_{ \bm{x} + \bm{\delta} }\right)
        \right],
    \label{eq.XYLocalInternalE}
\end{equation}
where $\bm{\delta}$ is a unit vector connecting the site $\bm{x}$ with one of its four nearest neighbours. In step 3, probabilistic decision for spin updating is made. If $p_{\bm{x}} < P(\Delta E_{\bm{x}})$, $\theta_{ \bm{x} }$ will update to $\theta'_{ \bm{x} }$. Otherwise, the spin direction will remain unchanged as $\theta_{ \bm{x} }$. 
\begin{figure}[h]
\includegraphics[width=0.47\textwidth]{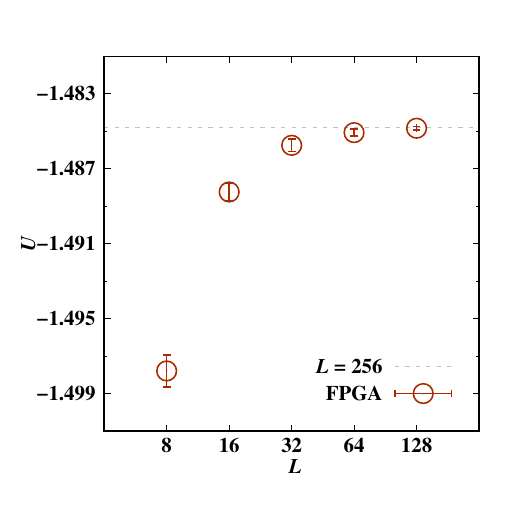}
\caption{The internal energy of a single spin in the XY model calculated on FPGA. The red hollow circles represent the numerical values of the internal energy for different $L$, with error bars indicating a 95$\%$ confidence interval. The dashed horizontal reference line represents the average internal energy per spin in a system with $L=256$ computed using the CPU.}
\label{fig:energyxy}
\end{figure}
After dividing the system into two sublattices, the three steps described above can be executed concurrently. This allows each site within one sublattice to evolve independently and in parallel because the system only has nearest neighbor interaction, and the rotation of a spin has no impact on the probability of rotation of another spin in the same sublattice.
%In other words, we changed the period of momentum space from $(2\pi, 2\pi)$ to $(\pi, \pi)$.

%To optimize the compatibility with the FPGA hardware framework, we separated the original state $\theta_{ \bm{x} }$ and the updated state $\theta'_{ \bm{x} }$ of the spin as input and output data streams, respectively. Furthermore, we also ensured the feasibility of block evolution for computing larger systems with limited computational resources.

We then calculate the internal energy of a single spin for $i$-th simulation on FPGA, defined as
\begin{equation}
    U_{i} = \frac{1}{L^{2}} \langle H_{\rm XY} \rangle
    \label{eq.XYE}
\end{equation}
where the $\langle \cdot \rangle$ denotes Monte Carlo steps average. Multiple simulations were performed with a series of sizes at temperature $T=0.85$. Each size had 10 different initial random seeds, resulting in 10 different trajectories in phase space. All simulations consisted of 1.2 million Monte Carlo (MC) steps. For each simulation, the last 1 million MC steps were used to measure the average internal energy $U_i$ of the corresponding trajectory, where the subscript $i$ denotes the $i$-th seed. The final result for the internal energy is calculated as $U = \frac{1}{10} \sum_{i=1}^{10} U_i$. The internal energy $U$ is plotted in Fig. \ref{fig:energyxy}. 
As $L$ increases from 8 to 128, the energy results obtained by FPGA calculations approach more closely to the CPU simulation results for $L=$256.

%As the system size raises up, the spin energy increases and tends to converge.

\begin{figure}[h]
\includegraphics[width=0.44\textwidth]{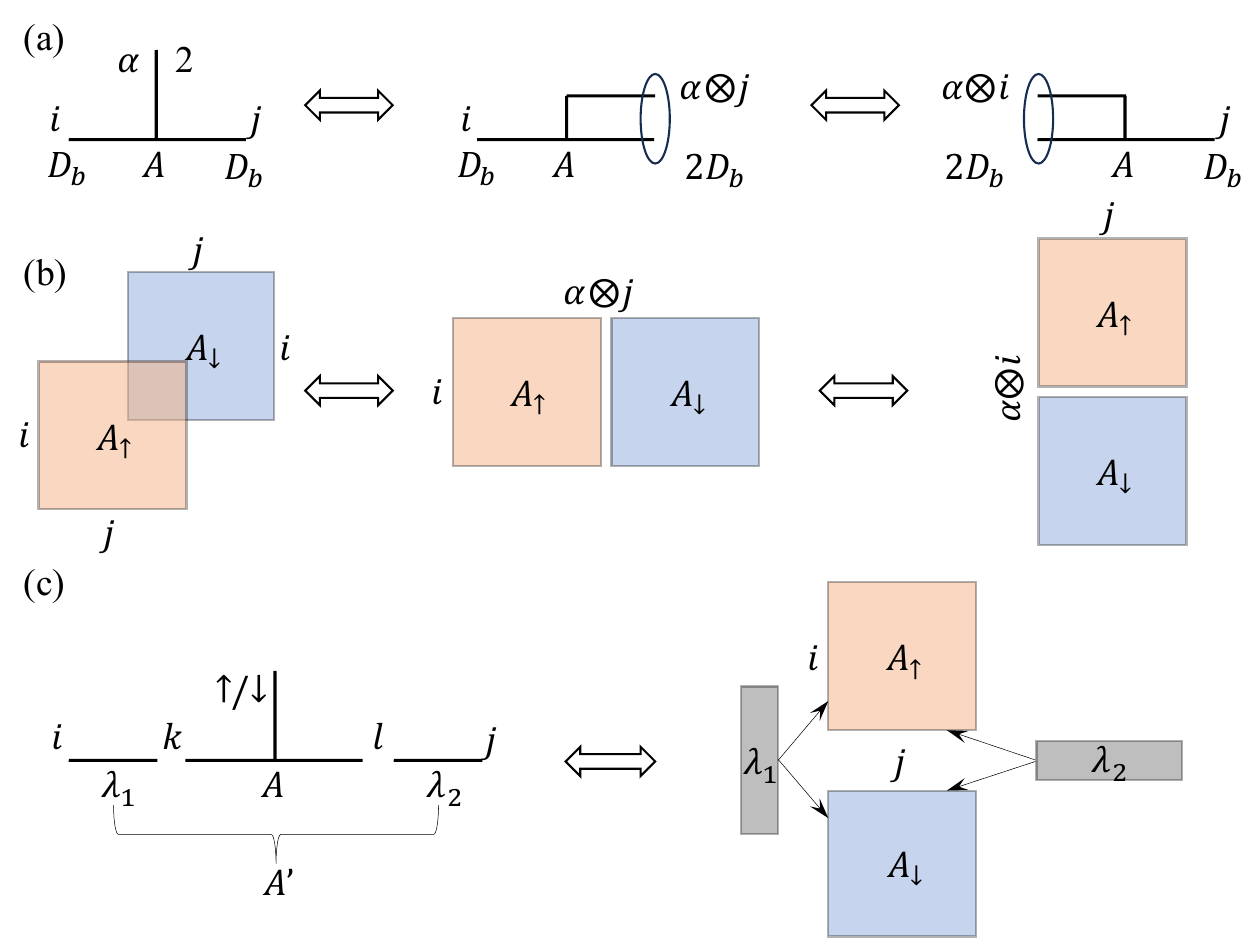}
\caption{The different representations of tensor reshaping operations. (a) demonstrates the transformation of a third-order tensor $A_{ij}^{\alpha}$ into two different forms of second-order matrices, (b) depicts the specific configurations of matrix elements corresponding to the transformations, and (c) illustrates the equivalence between the tensor contraction operation $A'=\lambda_1 A\lambda_2$ performed through reshaping on the left and the representation of matrix elements on the right.}
\label{fig:itebdmatrix}
\end{figure}

%Fig.~\ref{fig:itebdmatrix}
\section{Detail of the iTEBD calculation for the Heisenberg chain}
In the main text, we point out that tensor computations on FPGA need to avoid operations like permutation or reshaping to improve computational efficiency. Here, we need to clarify that the new form of tensor computations in FPGA is equivalent to the old form used in traditional tensor networks, which involves operations like permutation or reshaping. 
%Here, we illustrate this fact with a simple tensor contraction as an example.
In the following, we illustrate this fact by taking the process of contracting matrices $\lambda_{1,2}$ into the local tensor $A$ in the pre-SVD step as an example.
In traditional tensor networks, a contraction between a pair of tensors is recast to a matrix multiplication for generality and parallelism. This pair of tensors are reshaped to matrices where the tensor indices to be contracted (not contracted) are rearranged as the row (column) or column (row) indices. For instance, the local tensor $A$ has one spin index, $\alpha$, which can be $\uparrow$ or $\downarrow$, and two virtual indices, $i$ and $j$, whose dimensions are both $D_b$ [shown in Fig.~\ref{fig:itebdmatrix}(a)]. When $i$ is contracted with $\lambda_1$ at the left hand side, $A$ is reshaped to a $D_b \times 2 D_b$ matrix, where indices $\alpha$ and $j$ are combined into column index of the matrix. After $\lambda_1$ is contracted, $\alpha$ and $i$ are then combined into row index, to contract $j$ with $\lambda_2$ at the right hand side. Upon careful observation, one can find that tensor reshaping merely involves rearranging
the tensor elements, manifesting as fixed blocks of elements appearing at different positions [shown in Fig.~\ref{fig:itebdmatrix}(b)]. Therefore, we can operate on each block of elements $A_{\uparrow}$ or $A_{\downarrow}$ separately instead of operating on the entire tensor $A_{ij}^\alpha$. Considering the $\lambda$s are diagonal matrix, the contraction can be further simplified by taking $\lambda$s as vectors, like [shown in Fig.~\ref{fig:itebdmatrix}(c)]
\begin{equation}
A^{\prime}_{\uparrow/\downarrow,ij} = \sum_{kl} \delta_{ik} \lambda_{1,ik} A_{\uparrow/\downarrow,kl} \delta_{lj} \lambda_{2,lj} = \lambda_{1,i} A_{\uparrow/\downarrow,ij}\lambda_{2,j},
\end{equation}
where the new quantity $A'$ is the result of connecting $\lambda$s and $A$ together, $\delta$ is Kronecker delta. 
This operation for the tensor elements can be calculated parallelly. 
In the pre-svd step, other tensor network contraction can be operated similarly, shown in the Fig.~3 in the main text. 
\begin{figure}[h]
\includegraphics[width=0.35\textwidth]{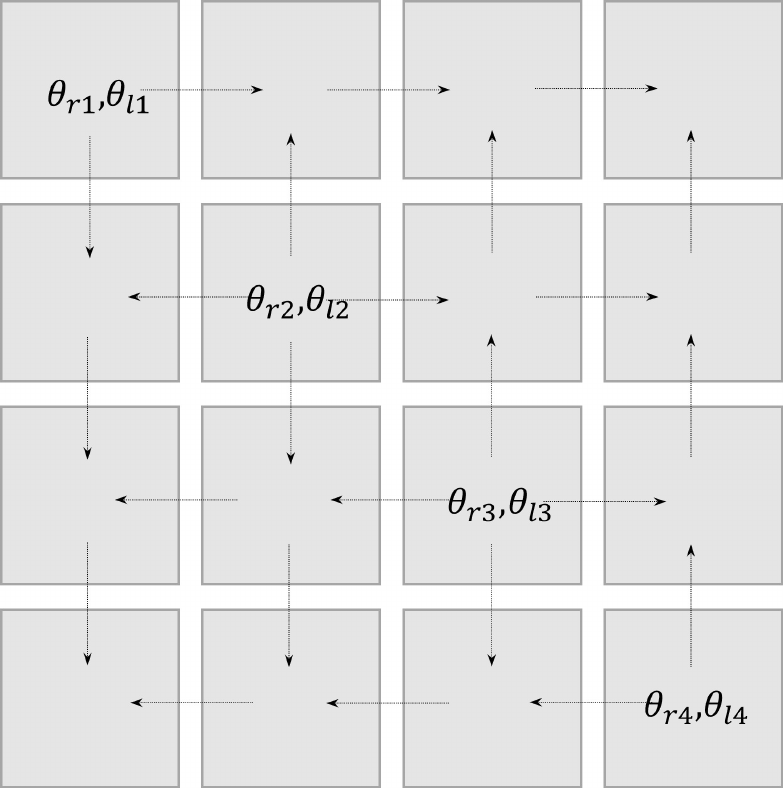}
\caption{Two-sided Jacobi rotation (illustrated using an $8\times 8$ matrix as an example), where each square represents a $2\times 2$ matrix. Diagonalization of the four $2\times 2$ matrices on the diagonal squares yields the corresponding left and right rotating angles $\theta_{li}$ and $\theta_{ri}$, which are then used to rotate the other $2\times 2$ matrices in the squares.}
\label{fig:jacobisvd}
\end{figure}

Next, in the svd step, one need to obtain the first $D_b$ singular values of a $2D_b\times 2D_b$ matrix $M$. We use two-sided Jacobi algorithm to perform the SVD calculation $\Lambda=U^\dagger MV$. Firstly, we separate the $2D_b\times 2D_b$ matrix $M$ into $D_b\times D_b$ small matrices with dimension $2\times 2$. This forms a $D_b\times D_b$ block matrix with row and column indice $(i,j)$. The matrix $U$ (or $V$) is then the products of left (or right) Jacobi rotation matrices $J_{r/l}$ obtained by diagonalizing the $2\times 2$ blocks in the diagonal direction. At each $i$, the $J_{r_i/l_i}$ has the form,
\begin{equation}
J_{r_i/l_i}=
\begin{bmatrix}
\cos\theta_{r_i/l_i} & \sin\theta_{r_i/l_i}\\
-\sin\theta_{r_i/l_i} & \cos\theta_{r_i/l_i}
\end{bmatrix}
\end{equation}  
The nondiagonal $2\times 2$ blocks with the row and column indice $(i,j)$ are rotated by the angles $\theta_{l_i}$ and $\theta_{r_j}$. Figure.~\ref{fig:jacobisvd} shows this procedure with an example of $D_b=4$. 

After one step of the rotation is performed, the matrix $M$ is transformed into a new matrix $M'$ to iterate the next rotation. The transformation is according to a systolic array~\cite{Luk1985s}, where the row/column indice of $M$ changes with the order
\begin{eqnarray}
\nonumber
2i-1&\rightarrow& 2i+1, \\\nonumber
2i+2&\rightarrow& 2i 
\end{eqnarray}
at $1\le i\le D_b-1$ except for the two boundary cases where $1\rightarrow 1$, $2\rightarrow 3$ and $2D_b-1\rightarrow 2D_b$. Take $D_b=4$ for example, start from the ordinary order from one to eight, this systolic array is

\begin{eqnarray*}
\nonumber
1,2,3,4,5,6,7,8;\\ \nonumber
1,4,2,6,3,8,5,7;\\ \nonumber
1,6,4,8,2,7,3,5;\\ \nonumber
1,8,6,7,4,5,2,3;\\ \nonumber
1,7,8,5,6,3,4,2;\\ \nonumber
1,5,7,3,8,2,6,4;\\ \nonumber
1,3,5,2,7,4,8,6; 
\end{eqnarray*}    

After seven steps, the array goes back to the ordinary order. By using this systolic array, all the possible combinations of the $2\times 2$ blocks are considered. Iterating these procedure will converge the Matrix $M$ into a diagonal form. The new $\lambda_2$ is the singular values of the matrix $M$. In the post-svd step, the new $A_{\uparrow/\downarrow}$ or $B_{\uparrow/\downarrow}$ can be calculated by contracting $\lambda_1^{-1}$ into the matrices $U$ and $V$. 
the iTEBD iteration is operated to $A_{\uparrow/\downarrow}$, $B_{\uparrow/\downarrow}$ and $\lambda_1$, $\lambda_2$ alternately, until the ground state $\ket{\Psi}$ is obtained.   

\begin{figure}[h]
\includegraphics[width=0.47\textwidth]{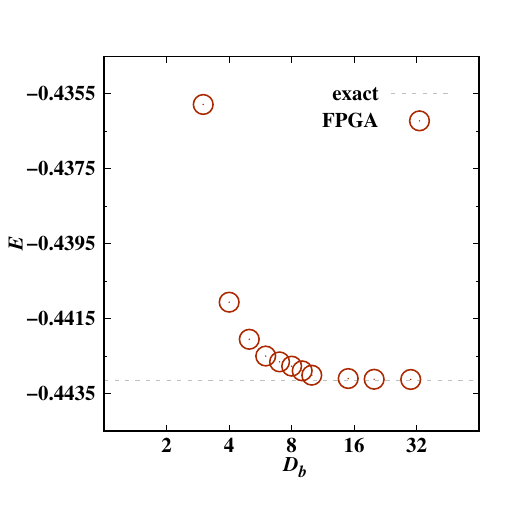}
\caption{The average energy of the Heisenberg chain computed from the ground state obtained on FPGA. The red hollow circles represent the energy values for different $D_b$. The dashed horizontal reference line represents the exact solution, $-\ln 2 + 0.25$.}
\label{fig:energyH}
\end{figure}

The ground state energy per bond for the Heisenberg chain is then calculated by 
\begin{equation}
    E = \bra{\Psi} H_{ij} \ket{\Psi}/\langle\Psi\ket{\Psi}
    \label{eq.HeisenbergE}
\end{equation}
The energy result at different $D_b$ from the ground state obtain from the FPGA is shown in Fig.~\ref{fig:energyH}. As $D_b$ is increased, it is converged into the exact result. 

\section{Performance comparison complement on GPU, CPU and FPGA}

\begin{figure}[h]
\includegraphics[width=1\textwidth]{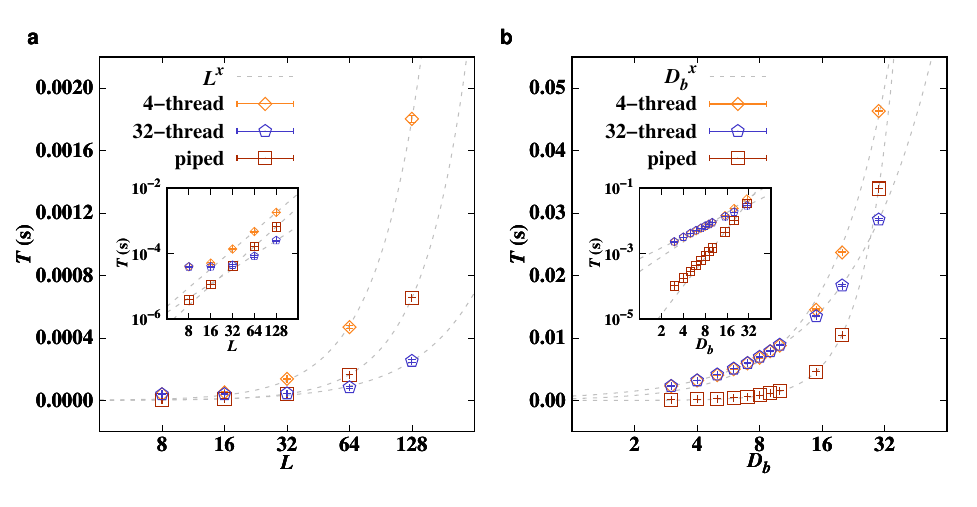}
\caption{The computational time per step of GPU with 4 active threads per warp compared with FPGA in pipelined parallel style for \textbf{a}, MC simulation of the XY model and \textbf{b}, iTEBD calculation of the Heisenberg chain. The red squares, the orange diamonds and the blue pentagons represent the computational time for FPGA in pipelined parallel style, GPU with 4 active threads per warp and GPU with 32 threads, respectively. The error bars indicate a 95\% confidence interval. The gray dashed lines represent the fitted results of the computational time with the fitting function $X^x$, where $x$ is the fitting parameters, and  $X = L$ or $X = D_b$ for \textbf{a} or \textbf{b}. The fitted results for $x$ for GPU with 4 active threads per warp is 1.91 and 1.49 in \textbf{a} and \textbf{b}, respectively.}
\label{fig:thread4time}
\end{figure}

The performance of FPGA and GPU with four threads per warp is further compared in Fig.~\ref{fig:thread4time}. The computation speed of FPGA can surpass that of GPU with four threads at moderate system size.
Additionally, by comparing the results of 4 threads and 32 threads, we observe that for small values of $L$ or $D_b$, a lower number of threads is sufficient, and increasing the thread count does not improve computational efficiency. Therefore, in Fig.~4, we only used larger $L$ or $D_b$ values for the scaling fit of the 32-thread GPU results. 
%The scaling behavior of GPU is better than that in the main text, demonstrating a feasible way with large potential for accelerating the many-body calculation which can also be applied to FPGA in the future.

The faster performance of GPU computation is mainly due to the higher number of threads. The parallelism of threads in a GPU can be equivalent to the unrolling parallelism in an FPGA. However, in an FPGA, unrolling consumes more hardware resources, such as memory or LUTs. In contrast, GPUs have ample memory. This comparison highlights that, given sufficient resources, an FPGA utilizing unrolling parallelism has enormous potential.

Regarding the multi-core performance on the CPU, we parallelized the MC simulation as a straightforward example. The parallel design for the square lattice proposed in this work is inherently adaptable to the multi-core execution and is implemented by OpenMP with different threads counts. Table \ref{tab:CPUMultiCore} illustrates the CPU speedup achieved by increasing the number of threads, with GPU results included for comparison. For both CPU and GPU, the computation time decreases with higher thread count, and the efficiency gains per additional thread diminish as the thread count becomes sufficiently large. However, the GPU exhibits superior and more nearly linear speedup with increasing thread counts compared to the CPU. This outcome suggests that one can focus on benchmarking the many-body parallel computation with GPU rather than CPU. 

\begin{table}[h!]
\centering
\caption{The speedup factor of the MC simulation at $L = 128$ with multi-threads execution on CPU and GPU. The speedup rate is represented by $t_1/t_{i}$, where $t_i$ is the computational time of one MC step with the number of threads $i$ }
\begin{tabularx}{0.5\textwidth}{C C C C C C}
\toprule
Rate & $t_1 / t_2$ & $t_1 / t_4$ & $t_1 / t_8$ & $t_1 / t_{16}$ & $t_1 / t_{32}$ \\
\midrule
CPU & 1.89 & 3.65 & 6.84 & 11.99 & 16.76 \\
GPU & 1.95 & 3.81 & 7.43 & 14.43 & 27.15 \\
\bottomrule
\end{tabularx}
\label{tab:CPUMultiCore}
\end{table}

\section{Details and comparisons of parallelization strategies applied to different lattice configurations}

\begin{figure}[h]
\includegraphics[width=1\textwidth]{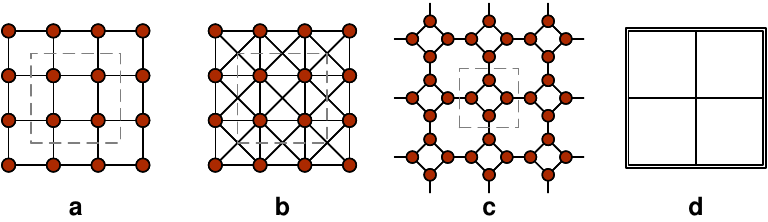}
\caption{Schematic diagram of different lattice configurations and partitions for \textbf{a}, square lattice, \textbf{b}, square lattice with next nearest neighbor interaction, \textbf{c} square octagonal lattice, and \textbf{d}, the supercell partition. The dark red circles, solid black lines, and dashed gray boxs represent the location of spins, the spin coupling, and the unit cell partition. The black double line box in \textbf{d} represents the whole lattice system, and is partitioned into four supercells. }
\label{fig:unitCell}
\end{figure}

The pipelined parallel strategy based on the bipartition of the system demonstrates effective speedup. While another way to further improve the efficiency of the parallel strategy is to restructure the partition of the system. As illustrated in the Fig.~\ref{fig:unitCell} \textbf{a}, the unit cell is enlarged and each spin in the same unit cell belongs to four different sublattices. The computation time of this quadripartition strategy for square lattice is illustrated in Fig.~5 \textbf{a} in the main text, which showcases superior computation speed than that of bipartition strategy. This quadripartition strategy can also parallelize the computation of another lattices, such as square lattice with next nearest neighbor (NNN) interaction and square octagonal lattice, as illustrated in Fig.~\ref{fig:unitCell} \textbf{b}, \textbf{c}. 
Under this quadripartition strategy, the unit cells are updated in the pipelined parallel arrangement. Although the latency for updating one unit cell is roughly stable, the number of unit cells traversed in one MC step increases in proportion to the number of spins $N$, implying the computation time scales as $O(N)$.

Compared to the square lattice with nearest neighbor (NN) interaction, incorporating NNN interaction doubles the computation time on the CPU. However, on the pipelined FPGA, the computation time for NNN interaction is less than twice that of NN interaction. The linear trend in Fig.~5 \textbf{a}, \textbf{b} and \textbf{c} reflects a similar behavior across different lattice configurations. Further investigation of the ratio between the intercept and slope in the fitted results suggests that the system's density does not significantly affect the scaling behavior or the computational time trend, provided that the system exhibits periodicity and appropriate optimization strategies are applied.

In order to compute sufficiently large system sizes, multiple unit cells reuse the same FPGA fabric blocks in pipelined arrangement. In other words, the number of unit cells that can be simultaneously updated is limited by the hardware resources which restricts further parallel improvement. However, in the small system sizes  we observe that, by investing enough hardware resources to allocate a unique fabric block to each spin in a sublattice for independent update, the MC step can be executed in a determinated latency which is independent from the spin numbers, demonstrating an $O(1)$ scaling. To be more specific, as illustrated in Fig.~\ref{fig:unitCell} \textbf{d}, the system is partitioned into four supercells where all the spins in the same sublattice are combined into one supercell. And the procedures for updating a supercell are fabricated into the FPGA, where four supercell update executions constitute one MC step. When the spin number increases, the supercell update still can be executed in an unchanged latency, if the unique update fabric block for each spin in the supercell is provided. This supercell strategy can perform $O(1)$ scaling behavior for all the three lattice configurations in Fig.~\ref{fig:unitCell}, implying that with sufficient resources, the reasonable utilization of system symmetry can keep the effect of denseness at a constant.

One priority of our parallel design is to strictly implement the Markov chain MC to FPGA. We actually calculated the value of probability $P(\Delta E_{x})$ with the same precision as CPU. It is reasonable to perform a slower computation speed for keeping the precision. The computation time for the convergence of Ising model on square octagonal lattice in our work is around $10^3$ microseconds, which is faster than CPU and close to the Graph-colored p-bit performance in Ref.~\cite{Chowdhury2023s}.

\section{Resource usage of FPGA in pipelined parallel design}

The scaling behavior of memory usage and computational cost can be theoretically analyzed from the algorithm structure. In the algorithm of MC simulation of XY model, each spin on the square lattice has its corresponding data, for example, the angle of the spin, and the updating procedure will do the same sampling and acceptance decision for all the spin. Thus the memory usage and computational cost for MC simulation is scaling as $O(L^2)$ which is equivalent to $O(N)$, where $L$ and $N = L^2$ are system size and number of spins in the system, respectively. As to the algorithm of iTEBD calculation of Heisenberg chain, the largest memory consumption is saving the one-gate updated but un-cut unit cell wave function which is a matrix with the dimension of $2D_b \times 2D_b$, implying that the scaling behavior of memory usage is $O(D_b^2 )$, and the singular value decomposition (SVD) of this matrix contributes the major computational cost of iTEBD calculation, which is scaling as $O(D_b^3)$. 

\begin{table}[h!]
\centering
\caption{The proportion of used computational resources to available resources of calculations with largest data sizes on FPGA in pipelined parallel style. The column of Data size lists the system size $L$ for the XY model, or the bond dimension $D_b$ for the Heisenberg chain. The available computational resources on the FPGA chip are listed in the FPGA hardware section in the main text.}
\begin{tabularx}{0.8\textwidth}{C C C C C}
\toprule
Data size	&BRAM\_18K (\%)	&DSP48E (\%)	&FF (\%)	&LUT (\%) \\
\midrule
$L=32$    &8.3   &37.7  &5.4   &17.9 \\
$L=64$    &14.4  &39.0  &5.6   &18.2 \\
$L=128$   &42.0  &39.0  &5.6   &18.2 \\
$D_b=15$  &25.1  &62.9  &19.6  &65.3 \\
$D_b=20$  &32.6  &63.2  &20.7  &66.8 \\
$D_b=30$  &42.5  &63.2  &24.3  &69.9 \\
\bottomrule
\end{tabularx}
\label{tab:FPGAresourceUsage}
\end{table}

The limited resources on the FPGA chip restrict our calculations in the moderate scale. We would like to illustrate the scalability of the proposed FPGA implementation for larger-scale computations from the perspective of computational resource usage. In Table.~\ref{tab:FPGAresourceUsage}, the computational resource usage percentage of the calculations with largest data sizes performed on the FPGA in pipelined parallel style is listed. With the growth of $L$ or $D_b$, the usage of BRAM\_18K significantly increases which is a natural result of storing more long-term data like spin angles or local tensors, while the usage of DSP48E, FF and LUT performs much slower growth, which means the FPGA still has sufficient capability to saving temporary data and doing logical or numerical computations. The distinct difference of the increasing tendency demonstrates that BRAM\_18K is the primary constraint which restricts the expansion of simulation or calculation scale. 

\begin{figure}[h]
\includegraphics[width=1\textwidth]{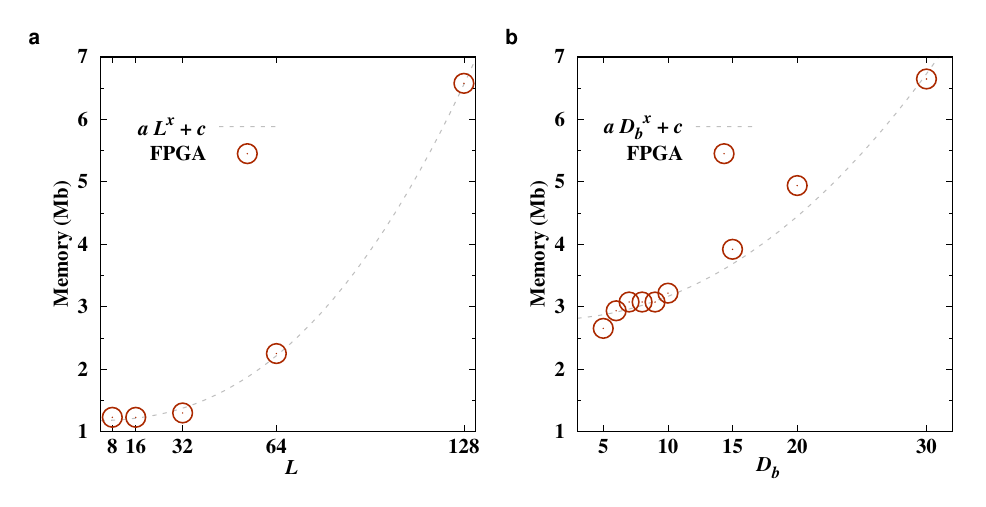}
\caption{The usage of BRAM\_18K converted to the form of equivalent memory size for \textbf{a}, MC simulations and \textbf{b}, iTEBD calculations performed on the FPGA in pipelined parallel style. The gray dashed line is the fitting function $aX^x+c$, where $x$ is the fitting parameter, and $X = L$ or $D_b$ for \textbf{a} or \textbf{b}. The fitted results for $x$ slightly larger than the theoretical value 2, which is 2.37 and 2.13 for \textbf{a} and \textbf{b}, respectively.}
\label{fig:memoryScale}
\end{figure}

To better illustrate the constraint of BRAM\_18K, the usage of BRAM\_18K for FPGA in parallel is plotted in Fig.~\ref{fig:memoryScale} in the form of equivalent memory size, which equals 15.6 Mb. 
%The utilization of BRAM\_18K deviates from the expected fit at small values of $L$ and $D_b$ due to the relatively large instance memory. Notably, for the data points not plotted and fitted, specifically at $L=8$ and $D_b = 3$, the proportion of instance memory is sufficiently large to render the deviation unsuitable for fitting.
%The BRAM\_18K usage of pipelined computations demonstrate a strong scaling behavior as %$L^{2.32}$ and $D_b^{1.35}$.
In the fitting process, we used the formula $aX^x + c$, where $X=L$ or $D_b$ and $c$ represents the inherent memory required for computation initialization, primarily stemming from Instance and Register.
This scaling behavior suggests that FPGA has deterministic requirements for memory capacity when enlarging the many-body system scale, which is much smaller than that of CPU and GPU as shown in the main text. 
This deterministic small memory usage of FPGA benefits from our design of reducing intermediate variables as much as possible, allowing us to effectively enlarge the many-body system scale by using the FPGAs with large BRAM, for example, the FPGA chips with model of XC7K480T which have 74.65 Mb embedded memory capacity. Or we can implement the external memory devices that play similar role with BRAM, for example, the Double Data Rate (DDR) memory. The FPGA fabric also allows for highly efficient data transfer between logic block and DDR. One can further design the BRAM as a buffer for DDR to perform higher efficiency.

\section{Operand bit width optimization trial}% for computational time and resource usage}

\begin{table}[h!]
\centering
\caption{The computational time and resource usage for MC simulations at $L = 32$ on FPGA in pipelined parallel style after operand bit width optimization. The data type named as ($m$, floating) refers to $m$-bit floating point type. }
\begin{tabularx}{0.9\textwidth}{C C C C C C}
\toprule
Data type &T ($\mu s$)	&BRAM\_18K (\%)	&DSP48E (\%)	&FF (\%)	&LUT (\%) \\
\midrule
(64, floating) &42.36 &8.3  &37.7  &5.4  &17.9 \\
(32, floating) &42.29 &7.0  &33.2  &4.8  &16.3 \\
(32, 16)       &41.82 &2.5  &10.0  &2.9  &23.1 \\
(24, 12)       &41.75 &2.5  &7.9   &1.8  &15.4 \\
(16, 8)        &41.68 &2.5  &6.9   &1.3  &9.6 \\
\bottomrule
\end{tabularx}
\label{tab:bitWidthTimeResource}
\end{table}

We optimized the operand bit width for the pipelined MC simulation at $L = 32$ on FPGA by converting the data type of spin angles from double-precision floating-point (64-bit) to single-precision floating-point (32-bit) and arbitrary precision fixed-point. Each fixed-point data type in this optimization is represented by a pair of numbers $(m, n)$, indicating a total bit-width of $m$ bits, with $n$ integer bits and $m - n$ fractional bits. The data type of random numbers remained unchanged to maintain the consistency of the sampling chain. And the computational time is evaluated by the latency of one MC step, where one latency cycle is equivalent to 10 nanoseconds.

As shown in Table~\ref{tab:bitWidthTimeResource}, the operand bit width optimization slightly improves the computational time per MC step and significantly reduces the usage of resources, such as BRAM\_18K and DSP48E. However, this optimization notably decreases the accuracy of the simulations as well. After converting the data type to fixed-point, the internal energy of a single spin becomes less precise in the percentile compared to the original double-precision result.

\section{Power efficiency comparison}

The FPGA showcased lower power usage than that of the CPU and GPU in this work. The FPGA-embedded board was powered by a power source with a constant voltage of 12 volts and a variable current ranging from 0.1 to 1 ampere, resulting in a power consumption range of 1.2 to 12 watts. Power usage of the CPU and GPU was obtained by querying their built-in sensors. The power consumption for representative calculations is listed in Table~\ref{tab:powerUsage}. Considering the GPU is working with a idle CPU with a power of 78 watts, both the CPU and GPU exhibited higher power usage than FPGA.

\begin{table}[h!]
\centering
\caption{The power consumption of the MC simulations at $L = 128$ and iTEBD calculations at $D_b = 30$ performed on FPGA, CPU and GPU.}
\begin{tabularx}{0.5\textwidth}{C C C C}
\toprule
Data size &CPU (W) & GPU (W) & FPGA (W) \\
\midrule
$L = 128$  &87     & 10      & 1.2 - 12\\
$D_b = 30$ &89     & 10      & 1.2 - 12\\
\bottomrule
\end{tabularx}
\label{tab:powerUsage}
\end{table}

\end{document}